\documentclass[sn-mathphys]{sn-jnl}% Math and Physical Sciences Reference Style
\jyear{2021}%

\theoremstyle{thmstyleone}%
\theoremstyle{thmstyletwo}%

\theoremstyle{thmstylethree}%
\usepackage{amsmath}
\DeclareMathOperator*{\argmin}{argmin} 
\DeclareMathOperator*{\argmax}{argmax}

\begin{document}

\title[Semiparametric Detection of Change-points in Volatility Dynamics]{A Semiparametric Approach to the Detection of Change-points in Volatility Dynamics of Financial Data}


\author[1]{\fnm{Huaiyu} \sur{Hu}}

\author*[1]{\fnm{Ashis} \sur{Gangopadhyay}}\email{ag@bu.edu}

\affil[1]{\orgdiv{Department of Mathematics and Statistics}, \orgname{Boston University}, \orgaddress{\street{111 Commington Mall}, \city{Boston}, \postcode{02215}, \state{MA}, \country{USA}}}

\abstract{One of the most important features of financial time series data is volatility. There are often structural changes in volatility over time, and an accurate estimation of the volatility of financial time series requires careful identification of change-points.  A common approach to modeling the volatility of time series data is based on the well-known GARCH model. Although the problem of change-point estimation of volatility dynamics derived from the GARCH model has been considered in the literature, these approaches rely on parametric assumptions of the conditional error distribution, which are often violated in financial time series. This may lead to inaccuracies of change-point detection resulting in unreliable GARCH volatility estimates. This paper introduces a novel change-point detection algorithm based on a semiparametric GARCH model.  The proposed method retains the structural advantages of the GARCH process while incorporating the flexibility of nonparametric conditional error distribution. The approach utilizes a penalized likelihood derived from a semiparametric GARCH model along with an efficient binary segmentation algorithm.  The results show that in terms of the change-point estimation and detection accuracy, the semiparametric method outperforms the commonly used Quasi-MLE (QMLE) and other variations of GARCH models in wide-ranging scenarios.}




\maketitle

\section{Introduction}

In time series, particularly in financial time series, it is very common to observe structural changes at various time points in the data.  These changes, known as change-points or breakpoints, separate data into distinct segments. The problem of change-point detection attracts significant attention in widespread industries, including finance \cite{dias2004dynamic}, genomics \cite{venkatraman2007faster}, geology \cite{gupta1996detecting}, climate data analysis \cite{reeves2007review}, audio analysis \cite{gillet2007correlation}, oceanography \cite{killick2010detection}, and many other fields. In this paper, we focus on the change-point detection problem in financial time series.  In particular, we consider the problem of identification of changes in volatility, a question of utmost importance in financial data \cite{chen1997testing}.

The change-point detection can be viewed as a model selection problem involving the trade-off between model complexity and the resulting performance of the model. Therefore, one general approach is to define a cost function for segmentation \cite{braun2000multiple, zhang2007modified,killick2012optimal}, such that both the number and positions of change-points are decided by minimizing the penalized \cite{lee1995estimating} or constrained cost function \cite{braun1998statistical}. In the change-point literature, a widely utilized cost function is twice the negative log-likelihood \cite{haynes2014efficient, chen2011parametric, fotopoulos2010exact}. Therefore, an accurate estimation of the likelihood function is crucial in change-point detection problems. In this paper, we focus on the problem of volatility estimation of financial data, and in that context, we introduce a new method of change-point detection via a semiparametric likelihood. 

During the last several decades, a significant body of knowledge has been developed in the area of modeling financial time series.  In particular, after the  introduction of  Autoregressive Conditional Heteroskedasticity (ARCH) model by  Engle \cite{engle1982autoregressive} and  the Generalized  ARCH  (GARCH) model by Bollerslev \cite{bollerslev1986generalized}, considerable research has been focused on modeling the volatility of observed financial returns. These models have been shown to capture the stylized features of financial data, such as volatility clustering and leptokurtosis. Therefore, despite certain limitations, these models have become essential instruments in understanding the volatility of financial data.

The GARCH(p, q) model for the volatility of returns $y_t$ assumes the form:
\begin{equation}
	y_t = \sigma_t\epsilon_t, \quad \sigma_t^2 = \omega + \sum_{i=1}^{p}\alpha_iy^2_{t-i}+\sum_{j=1}^{q}\beta_j\sigma^2_{t-j} \label{GARCHpq}
\end{equation}
where $\omega \ge 0, \alpha_i \ge 0, \beta_j \ge 0$ and $\sum \alpha_i + \sum \beta_j < 1$. The white noise process $\epsilon_t \stackrel{i.i.d}{\sim} f$ assumes $E(\epsilon_t) = 0$ and $Var(\epsilon_t) = 1$ . In this model, $y_t$ is the centralized values of the return $r_t$, i.e., $y_t=r_t-\mu_t$, where $\mu_t$ stands for a smooth trend, and $\sigma_t$ is a stochastic process known as volatility and assumed to be independent of $\epsilon_t$.

When the conditional error distribution $f$ is known, parameters $\theta = (\omega, \alpha_1, ..., \alpha_p, \beta_1, ..., \beta_q)$ in the GARCH model are estimated directly by maximum likelihood estimation (MLE). However, in practice, the error distribution is unknown, and the usual approach is to assume a parametric form for $f$, which leads to an estimation method called quasi-maximum likelihood estimation (QMLE). In earlier literature, the conditional error distribution is generally assumed to be Gaussian \cite{weiss1986asymptotic, newey1997asymptotic}. The GARCH parameter estimates of the Gaussian QMLE are consistent regardless of the true error distribution \cite{weiss1986asymptotic}. It is a common approach to the estimation of GARCH parameters.  For example, in most R packages, such as the fGarch package \cite{wuertz2013package}, Gaussian QMLE is applied by default to estimate parameters $\theta$.

However, for financial data, the conditional normality assumption of $y_t$ is usually violated, and the specification of an appropriate parametric error distribution is often difficult \cite{Diebold1988Empirical}. In addition, Gaussian QMLE suffers from efficiency loss in cases where the true error distribution is non-Gaussian, such as Student's t, generalized Gaussian and other heavy-tailed densities \cite{engle1991semiparametric}. Hall and Yao \cite{hall2003inference} showed that the asymptotic normality and convergence rates QMLE are incorrect when the error distribution is heavy-tailed.  A possible solution to the problem is to consider a semiparametric estimator of the GARCH parameters, where assumption of the error distribution is relaxed by taking it to be any absolutely continuous pdf $f$, and the likelihood is constructed by estimating $f$, possibly by utilizing the residuals based on an initial estimate of the model parameters.  Some approaches in this vein include discrete maximum penalized likelihood estimator (DMPLE) \cite{engle1991semiparametric} and MLE derived from a likelihood based on a nonparametric density estimate (SMLE) \cite{di2011efficiency, di2013one}. A review of the current state of the semiparametric estimation of the GARCH model is given by Di and Gangopadhyay \cite{digangop2011review}.

In this paper, we argue that the successful identification of change-point of volatility for financial data can be achieved via a semiparametric likelihood, and we show that the performance of the proposed method is superior to parametric likelihood under model miss-specification. Even when the likelihood is correctly specified, which is an oxymoron in real applications, the performance of the semiparametric method is at par with its parametric counterpart.  A key step of the analysis is to develop a semiparametric likelihood, and we have utilized a one-step likelihood function of the GARCH model introduced in \cite{di2013one}. The term ``one-step" refers to fact that the likelihood is evaluated in a single step via a search on the parameter space, as opposed to a two-step method that requires an initial estimate of the model parameters to generate the residuals, which in turn is used to nonparametrically estimate the error density leading to the likelihood function \cite{di2011efficiency}. The method is discussed in detail in the next section. The one-step SMLE is consistent and asymptotically Gaussian \cite{di2013one}. The likelihood function in this model is untrimmed, which implies no information is abandoned in the estimating procedure. Simulation results show that the one-step SMLE approach is more robust with smaller bias and variability of GARCH estimators than the two-step SMLE and QMLE \cite{di2013one}. 

The paper address the question of identification and estimation of volatility change-points in financial data.  This is a fundamental question in volatility estimation of financial time series as financial data observed over a long period is likely to have structural changes. The traditional method of volatility estimation based on a single GARCH(1,1)  process will not result in a reliable estimate of volatility.  Only a few papers in the literature have attempted to address this question, and as far as we know, all these works are centered around Gaussian GARCH(1,1) process. Therefore, in this paper, we address the critical question based on a novel approach to change-point detection of volatility based on the one-step SMLE-GARCH model developed in \cite{di2013one}.  We propose an efficient algorithm to search for change-points based on Binary Segmentation.  Since a theoretical analysis of the problem is intractable, we provide the results of extensive simulations that show the superiority and reliability of the semiparametric algorithm compared to the parametric models. We believe that the work presented here makes a significant contribution to the realm of volatility estimation of financial data.

The rest of this paper is organized as follows. In Section \ref{Section2}, we introduce the one-step semiparametric GARCH estimator. In Section \ref{Section3}, we review change-point detection algorithm based on the semiparametric likelihood and describe a binary segmentation approach of model estimation. Section \ref{Section4} presents the simulation results of the performance of proposed semiparametric change-point detection method and applications to the financial time series. Discussions and conclusions are in Section \ref{Section5}.

\section{Semiparametric estimation in GARCH Model}\label{Section2}
In the statistics literature, the term \textit{semiparametric} has been used to refer to different estimation scenarios. One commonly studied problem is where certain components of a model are characterized parametrically, while other components are described nonparametrically \cite{Ruppert2003}. In GARCH models, this idea is often incorporated by having a nonparametric functional form for the volatility function \cite{pagan1990, Hardle97, Linton05, Yan.06}.  However, in the present work, we focus on the scenario involving a fully parametrically specified model and an unknown innovation pdf. In this section, we describe the one-step semiparametric GARCH estimation introduced by Di and Gangopadhyay \cite{di2013one}, which is the foundation of the semiparametric change-point detection method proposed in the next section.

Let's denote the parameters of GARCH model as $\theta = (\omega, \alpha_1, ..., \alpha_p, \beta_1, ..., \beta_q)$, and let the observed time series is given by  $y = (y_1, ..., y_n)$. Based on GARCH model in Equation (\ref{GARCHpq}), the residuals are $\epsilon_t(\theta) = y_t/\sigma_t(\theta)$, where $\sigma_t(\theta)$ is the volatility at time $t$ derived by $\theta$. The true log-likelihood of this GARCH model is
\begin{equation}
	L_n(\theta) = \frac{1}{n}\sum_{t=1}^n l_t(\theta) = \frac{1}{n}\sum_{t=1}^n log[\frac{1}{\sigma_t(\theta)} f(\frac{y_t}{\sigma_t(\theta)})],\label{origin}
\end{equation}
where $f$ is uniformly bounded and continuous pdf. Since in practice $f$ is unknown, we replace it with a nonparametric kernel density estimate $\hat f_n$ given by
\begin{equation}
	\hat f_n(z) = \frac{1}{n h_n}\sum_{t=1}^nK(\frac{z-\epsilon_t(\theta_n)}{h_n}), \label{kernel}
\end{equation}
where $K(\cdot)$ is a regular density kernel and $h_n$ represents the bandwidth, or the smoothing parameter. In the context of nonparametric density estimation, an important question is the choice of the bandwidth $h_n$. There are many data-dependent choices of the bandwidth, including $nrd0$ \cite{silverman1986density}, $nrd$, $ucv$, $bcv$ \cite{scott2015multivariate}, and $SJ$ \cite{sheather1991reliable}. In particular, the so-called normal reference bandwidths $nrd0$ and $nrd$ are computed based on the prior assumption that the true density $f$ is Gaussian, and these approaches are probably the easiest and fastest to implement. After comparing the simulation performances utilizing bandwidth selection methods $nrd0$, $nrd$, $ucv$, $bcv$ and $SJ$, we observed no obvious differences in the results. Thus, due to the computational efficiency, in this paper we utilize $nrd$, also known as the rule-of-thumb bandwidth selection method \cite{venables2002modern}.

Replacing the true density $f$ by its estimate $\hat f_n$ in Equation (\ref{origin}), the corresponding semiparametric log-likelihood function at any given $\theta$ is 
\begin{equation}
	\hat L_n(\theta) = \frac{1}{n}\sum_{t=1}^n \hat l_t(\theta) = \frac{1}{n}\sum_{t=1}^n log[\frac{1}{\sigma_t(\theta)}\hat f_n(\frac{y_t}{\sigma_t(\theta)})].\label{onestepllh}
\end{equation}
Therefore, the proposed semiparametric estimator of $\theta$ (SMLE) is achieved by 
\begin{equation}
	\hat \theta_n^{SMLE} = \argmax_{\theta \in \Theta}\hat L_n(\theta).\label{llh}
\end{equation}

Note that the QMLE is a special case of the SMLE in the sense that the fixed parametric form for $f$ assumed in the QMLE is replaced by an estimate $\hat{f}_n$ in the SMLE. Therefore, SMLE is intuitively better than the QMLE since a suitable choice of $\hat{f}_n$ may converge to $f$ in some sense, whereas an arbitrary parametric choice of $f$ does not. Hence, with a large sample size (which is generally the case in the financial time series), the semiparametric likelihood based on $\hat{f}_n$ can better reflect the true likelihood, compared to a quasi-likelihood based on a parametric assumption on $f$.

The properties and applications of such semiparametric estimators have been investigated in the literature quite extensively.  Engle and Gonzalez-Rivera \cite{engle1991semiparametric} have discussed an application of the semiparametric volatility estimate of \pounds/\$ exchange rate returns. Since the return series exhibits a clear peak and heavy tail, the validity of the assumption of unconditional Gaussian or unconditional $t-distribution$ is questionable. They showed that the semiparametric approach is more appropriate in representing the distribution of the return series. Drost and Klaassen \cite{Dro.97} studied the exchange rates of a total of fifteen currencies on the U.S. dollar under their adaptive estimation setup. They compared the semiparametric approach to the QMLE via bootstrap and reported that the semiparametric estimators have smaller estimation standard error. These studies illustrate the practical benefits of semiparametric estimation of the GARCH parameters. Therefore, semiparametric likelihood offers an appropriate and promising avenue for the identification and estimation of the change-points of volatility in financial data.

There are a couple of important issues related to SMLE worth mentioning. First, in the one-step estimation procedure, the residuals are not created from a predefined value or estimate of the parameters. Instead, the parameter being estimated is utilized directly to generate the residuals. In other words, the pdf $f$ in Equation (\ref{onestepllh}) is estimated from $\hat{\epsilon}_1(\theta),\hat{\epsilon}_2(\theta)....,\hat{\epsilon}_n(\theta)$. As a result, in the one-step procedure, $\hat{f}$ becomes a function of $\theta$, and hence it is a part of the likelihood to be maximized. Therefore, the procedure does not require an initial estimate of $\theta$.   The second issue is related to the parameter identifiability of the GARCH model.  In particular, the one-step SMLE approach introduces a parameter identifiability problem. Recall that in Equation (\ref{GARCHpq}), we assume $Var(\epsilon_t) = 1$. This condition is necessary to ensure the joint identifiability of $\theta$ and the overall scale of the GARCH model. In the SMLE method, an extra step is needed to ensure that the maximization is carried out within the parameter space that reflects the unit variance of $\epsilon_t$, i.e., $\Theta= \left\lbrace \theta : Var(\epsilon_t(\theta))=1\right\rbrace$. In our algorithm, the parameter identifiability is achieved by standardizing the residuals by subtracting the mean and dividing by the standard deviation before it is utilized to derive the kernel estimate $\hat{f}_n$. The standardization step simply ensures that the model residuals have zero mean and unit variance, and as discussed in \cite{di2013one}, this is a crucial step in the model estimation via semiparametric likelihood. The pseudocode of this one-step MLE of the semiparametric GARCH model is given in Algorithm \ref{semiparametric}. 

\begin{algorithm}
	\caption{One-step Semiparametric GARCH Model}\label{semiparametric}
	\begin{algorithmic}[1]
		\Require Data set $(y_1, y_2, ..., y_n)$
		\State Initialize the $\sigma_1(\theta) = std(y)$
		\State Compute $\sigma_t$ recursively according to Equation [\ref{GARCHpq}]
		\State Calculate the residuals $\hat \epsilon_t(\theta) = y_t/\hat \sigma_t(\theta)$
		\State Standardized the residuals as $\hat \epsilon_t(\theta)^* = \frac{\hat \epsilon_t(\theta) - mean(\hat \epsilon_t(\theta))}{std(\hat \epsilon_t(\theta))}$
		\State Estimate the density $\hat f_n $ with $\hat \epsilon_t(\theta)^*$ by Equation [\ref{kernel}]
		\State Estimate $\hat \theta$ by optimizing Equation [\ref{llh}] after plugging in $\hat f_n$ for $f$
	\end{algorithmic}
\end{algorithm}

In the next two sections, we will develop the algorithms for change-point estimation utilizing SMLE. We argue that the improved properties of  SMLE compared to the parametric methods in GARCH estimation translates to the superior performance of the proposed change-point estimation and detection algorithms.

\section{Semiparametric change-point models} \label{Section3}
This section introduces a change-point detection algorithm utilizing semiparametric likelihood and a search algorithm using binary segmentation.  As mentioned in the last section, the semiparametric  GARCH volatility estimation is a robust alternative to the parametric approaches, such as the Gaussian GARCH based method.  This section argues that by leveraging the semiparametric GARCH, the proposed volatility change-point detection and estimation algorithm provides a superior alternative to the parametric GARCH algorithms currently available in the literature. \par

\subsection{Model definition}
Given a sequence of observations ordered by time, $y_{1:n} = (y_1, ..., y_n)$, we denote data segment from $s$ to $t$ as $y_{s:t} = (y_s, ..., y_t)$ where $s \le t$. If there are $k$ change-points in the data set, these break points will divide the whole data set into $k+1$ segments. The $j$th segment of the data set includes observed samples $y_{\tau_{j-1}+1}, ..., y_{\tau_j}$. The notation of the location of the $j$th change-point is set as $\tau_j$ for $j = 1, ..., k$. In particular, $\tau_0 = 0$ and $\tau_{k+1} = n$.  The set of change-points is defined as $\tau = (\tau_0, ..., \tau_{k+1})$.\par

One prevalent approach to determine change-points is to minimize the sum of cost function across all segments. The cost of a segment $\tau_j$ is denoted as  $C(y_{(\tau_j + 1) : \tau_{j+1}})$. The change-points $\tau$ are estimated by:
\begin{equation}
	\hat \tau = \argmin_ \tau C_{k, n}=\argmin _\tau \sum^k_{j=0}C(y_{(\tau_j + 1) : \tau_{j+1}}) \label{cpdetect}
\end{equation}

The sum of cost functions $C_{k, n}$ monotonically decreases when the number of change-points $k$ increases. However, as $k$ increases, while the model is more flexible, it leads to the possibility of overfitting. To address this problem, penalized optimization has been used extensively to control the number of change-points \cite{lee1995estimating, killick2012optimal, maidstone2017optimal}. Thus, the change-points are identified by minimizing a penalized cost function given by
\begin{equation}
	\hat \tau = \argmin_ \tau C_{k, n} + \beta f(k)=\argmin _\tau \sum^k_{j=0}C(y_{(\tau_j + 1) : \tau_{j+1}}) + \beta f(k). \label{pencpdetect}
\end{equation}

In this paper, we have utilized -2$\times$log-likelihood as the cost function, where the likelihood is derived from the semiparametric approach outlined in the previous section. The choice of the penalty term controls the trade-off between the model performance and the associated cost,  which affects the segmentation results considerably. In the change-point literature, the most common technique is to apply a linear penalty in the number of change-points \cite{killick2012optimal, haynes2014efficient}, which is $\beta f(k) = \beta k$, for some constant $\beta>0$. In terms of choices of $\beta$, some of the common selections include Akaike's Information Criterion (AIC \cite{akaike1974new}, $\beta = 2p$), Schwarz Information Criterion (SIC \cite{schwarz1978estimating}, $\beta = p\,log(n)$) and $\beta = 2p\,log\,log (n)$ \cite{hannan1979determination}, where $p$ is the number of additional parameters introduced by increasing the number of change-points by 1. After comparing the results of change-point detection, we find that SIC is a better stopping rule, detecting a more correct number and positions of change-points compared to other penalty terms, such as AIC. Therefore, in multiple change-point detection algorithm, we choose SIC as the penalty term. For the sake of brevity, we will omit the results of comparison between different penalty terms.

\subsection{Binary segmentation in change-point detection}
Change-point literature mainly focuses on two classes of algorithms to solve Equation (\ref{pencpdetect}). The first is the dynamic programming algorithm depending on different pruning techniques, such as Pruned Exact Linear Time (PELT) \cite{killick2012optimal} and Pruned DP algorithm (pDPA) \cite{rigaill2010pruned}. However, the other much faster algorithm is the Binary Segmentation method, which is utilized in this paper.

Binary Segmentation is a generic search algorithm for multiple change-point detection proposed by Scott and Knott \cite{scott1974cluster}. The advantages of this algorithm include low computational complexity ($\mathcal{O}(n\log{}n)$, where $n$ is the number of data points), straightforward implementation and interpretation simplicity. This algorithm is widely used in change-point detection \cite{killick2012optimal, chen2011consistent, fryzlewicz2014wild} with accurate simulation results. Vostrikova \cite{vostrikova1981detecting} shows the consistency of binary segmentation approach for estimating the true change-point locations under suitable regularity conditions.

The algorithm is based on a single change-point detection. Once a change-point is identified, the data is split into two subsegments. The procedure is repeated until no additional change-points are detected. Full details for implementation are given in Algorithm \ref{bsalgorithm}. Binary Segmentation can be regarded as a regularized model selection method to find a solution to Equation (\ref{pencpdetect}). Each recursion of the algorithm introduces an additional change-point if and only if the new change-point induces a reduction in the total cost in Equation (\ref{pencpdetect}). Specifically, given any data segment $y_{s:t} = (y_s, y_{s+1}, ..., y_t)$, there exists one change-point $\tau$ if it satisfies
\begin{equation}
	C(y_{s:\tau})
	+ C(y_{(\tau+1):t}) + \beta f(k) < C(y_{s:t}), \label{bs}
\end{equation}
where, as mentioned earlier, the cost function is twice the negative log-likelihood of the selected time series data.

\begin{algorithm}
	\caption{Generic Binary Segmentation Algorithm}\label{bsalgorithm}
	\begin{algorithmic}[1]
		\Require Data set $(y_1, y_2, ..., y_n)$ \newline 
		A test statistic $\Lambda(\cdot)$ dependent on the data \newline
		An estimator of change-point position $\hat\tau(\cdot)$\newline
		A rejection threshold $RT$\newline
		Step length \newline
		\State Let $CP = \emptyset$, and $S = \{[1,n]\}$	
		\While{$S \neq \emptyset$}
		\State Choose an element of $S$, denote this element as $[s, t]$
		\If{$\Lambda(y_{s:t}) < RT$}
		\State remove $[s, t]$ from $S$
		\Else 
		\State remove $[s, t]$ from $S$
		\State calculate $r = \hat \tau(y_{s:t}) + s - 1$, and add $r$ to $CP$
		\If{$r - s > $step length}
		\State add $[s, r]$ to $S$
		\EndIf
		\If{$t - r > $step length}
		\State add $[r+1, t]$ to $S$.
		\EndIf
		\EndIf
		\EndWhile
		\State Return the set of change-points recorded $CP$ as output
	\end{algorithmic}
\end{algorithm}

The algorithm requires inputs that include a test statistic $\Lambda(\cdot)$, estimator of change-point position $\hat \tau(\cdot)$, and rejection threshold $RT$ (the same as penalty term $\beta f(k)$ in Equation (\ref{pencpdetect})). According to Equation (\ref{bs}), the test statistic $\Lambda(\cdot)$ is the distance between cost functions without and with change-point, which is $C(y_{s:t}) - (C(y_{s:\tau}) + C(y_{(\tau+1):t}))$, and the rejection threshold $RT$ acts as the stopping rule. A change-point is identified in the segment $y_{s:t}$ if $\Lambda(y_{s:t}) > RT$, then the position of change-point $r$ is stored as $\hat \tau(y_{s:t}) + s - 1$. The algorithm creates two sets of values, namely, $CP$ and $S$, where $CP$ is the set of detected change-points and  $S$ is the set of data segments that need to be checked for the identification of additional change-points. To improve the computational efficiency, in the algorithm, we have added a hyperparameter \textit{step length}. Instead of looking through every data point, we explore data segments only when the distance is large enough. We choose the \textit{step length} depending on the sample size, and it is updated within the algorithm as it searches through subsets of the data.  Specifically, if the length of a split data segment is less than the step length, it is unnecessary to split the segment and check change-point in this segment. In other words, in the segment $[s, t]$, if the length between interval bounds $s$ or $t$ and detected change-point $r$ ($s$ and $r$  or $t$ and $r$) is greater than the step length, new segments $[s, r]$ and $[r+1, t]$ will be added into the set $S$.

We will apply the Binary Segmentation algorithm to locate multiple change-points in Section \ref{multicp} and compare the performance of one-step SMLE with Gaussian QMLE in the volatility estimation in GARCH model. In Section \ref{application}, the algorithm is utilized to identify the change-points in the volatility of stock market indices, such as S \& P 500 and Dow Jones Industrial Average.

\section{Simulations and applications}\label{Section4}
In this section, we discuss the results of simulation studies to compare the methods based on Gaussian QMLE and SMLE in detecting change-points in GARCH volatility. To keep the discussion focused and concise, we will concentrate on the GARCH (1,1) model, which is also the most common GARCH process used in modeling volatility in financial data.  However, the results presented here can be generalized to any GARCH(p,q) model. Thus, the volatility equation is given by $\sigma_t^2 = \omega + \alpha y^2_{t-1}+\beta \sigma^2_{t-1}, t= 1,..., n$. When $\alpha\geq0$, $\beta\geq0$, and $\alpha + \beta < 1$, the process $y_t$ is covariance stationary.

There are two primary objectives of the simulation study. The first is to judge the performance of the methods in identifying the locations of change-points, and the other is to determine the correct number of change-point $k$. In Section \ref{singlecp}, we will consider a single change-point detection problem, and in Section \ref{multicp}, we will discuss the generalization of this method in the context of multiple change-points. In Section \ref{asymcp}, we will introduce the simulation results related to the performance of the semiparametric approach compared to the two of the most common asymmetric GARCH, namely, EGARCH and GJR-GARCH processes.  In Section \ref{application}, we will conclude with an application of the proposed semiparametric method in identifying the change points in Dow-Jones and S\&P500 data.

\subsection{Estimation of a single change-point}\label{singlecp}
We will first consider the data generating process (DGP) with one change-point. In particular, we will generate a random sample $(y_1,y_2,....,y_{t_{0}})$ from a GARCH(1,1) process with parameters $(\omega_1, \alpha_1,\beta_1)$ (DGP 1) and a random sample $(y_{t_{0}+1}, y_{t_{0}+2},....,y_n)$ from another GARCH(1,1) process with parameters $(\omega_2, \alpha_2,\beta_2)$ (DGP 2), where $t_0$ is the unique change-point. The parameters of two DGPs are considered with small $\alpha$ and large $\beta$, which is commonly observed in financial data, under the assumption of  $\alpha\geq0$, $\beta\geq0$, and $\alpha + \beta < 1$. Specifically, the DGPs in this single change-point detection process are designed as:
$$\textbf{DGP 1} \ GARCH(1,1): \omega_1 = 0.1, \alpha_1 = 0.05, \beta_1 = 0.9$$ 
$$\textbf{DGP 2} \ GARCH(1,1): \omega_2 = 0.15, \alpha_2 = 0.2, \beta_2 = 0.7$$

For this particular simulation, our primary objective is to establish the efficacy of our proposed semiparametric approach to change-point estimation and to compare its performance to Gaussian QMLE. To keep the discussion focused, in this subsection we fix the number of change-points to its true value 1, i.e., $k=1$, and we will consider the general problem of an unknown number of change-points in the next section. Therefore, the estimation is carried out by Equation (\ref{cpdetect}), and no penalty term is involved. As mentioned earlier, for simplicity and efficiency of computation, in SMLE we choose the smoothing parameter $h_n$  in Equation (\ref{kernel}) as the normal reference bandwidth $nrd$. For brevity, we omit the simulation results based on different optimal bandwidth selection approaches. However, the choice of a bandwidth selection method has almost no effect on the results presented here. 

\begin{figure}
	\centering
	\includegraphics[width=1\textwidth]{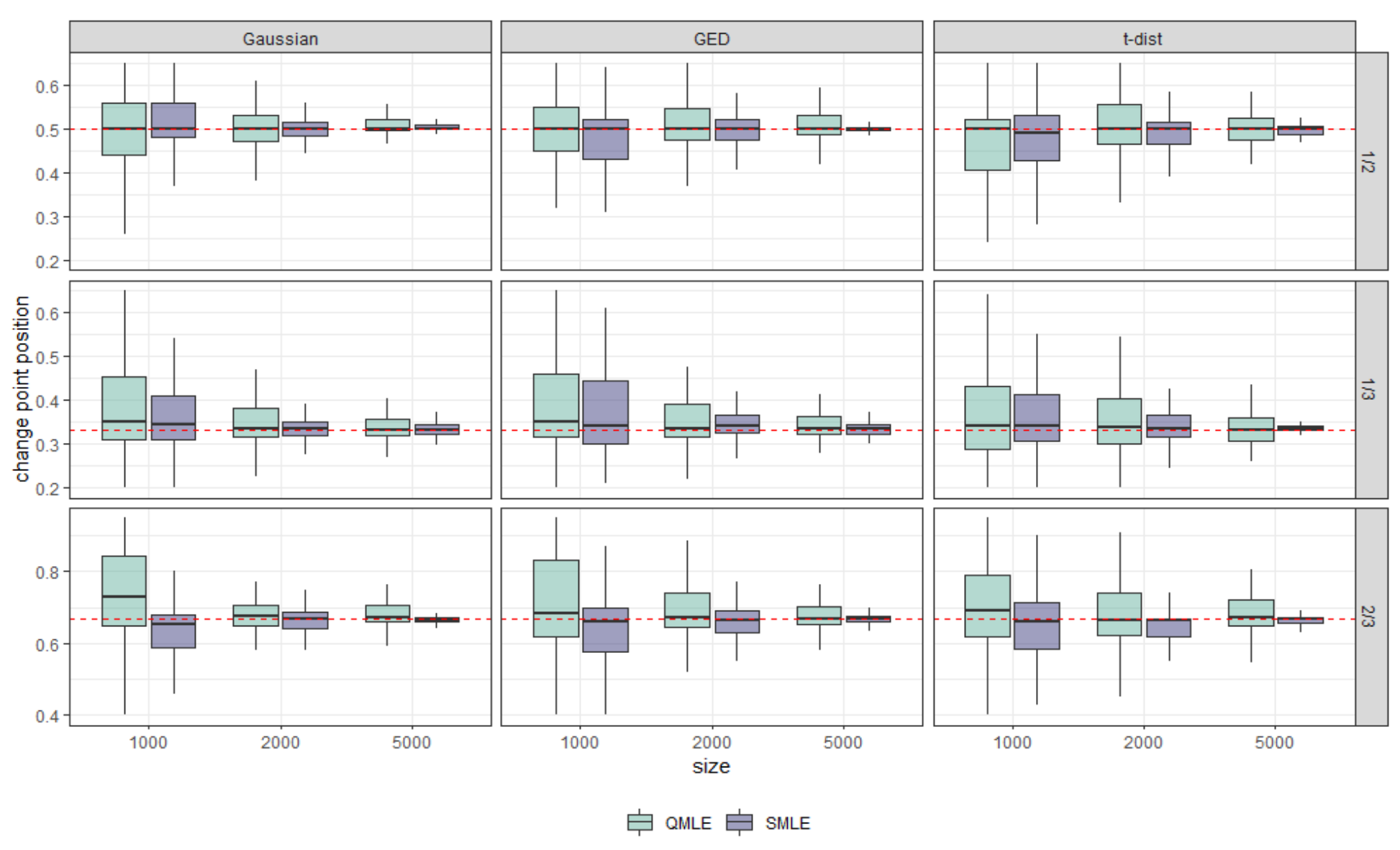}
	\caption{Boxplots of single change-point detection for sample sizes (1000, 2000, 5000), change-point positions (1/2, 1/3, 2/3), and error distributions standard Gaussian, GED with shape parameter equal to 1.5 (location and scale parameters are 0 and 1, respectively), and t-distribution with 6 degree of freedoms. The dotted line indicates the position of the true change-point. Each boxplot is based on 100 simulations.} \label{singlecp_plot} 
\end{figure}

The variables that may affect the performance of the change-point detection procedures include the length of the time series, the position of change-point in the time series, and the true error distribution $f$. To carefully explore all these variables, in the simulation, we have considered varying data lengths $n = (1000, 2000, 5000)$, and selected position of change-point at $q=(1/2, 1/3, 2/3)$ fraction of the length of the time series $n$. For example, a sample size $n = 1000$ and position $1/2$ represents the location of true change-point at $1000 \times 1/2 = 500$ in the simulated time series.  In terms of the true error density $f$, we have considered standardized Gaussian distribution $N(0, 1)$ and fat-tailed distributions, such as standardized Student’s t-distribution with the degree of freedom $6$, and Generalized Error Distribution (GED) with the shape parameter $1.5$ \cite{wuertz2013package}. This heavy-tailed distribution GED, also called generalized normal distribution, is commonly utilized as the assumed parametric error distribution $f$ in the ARCH(p) process. Suppose that $x_{\nu}$ follows a GED with $\nu$ degrees of freedom (shape parameter), then it has a probability density function given by:
$$f(x) = \frac{\nu\exp(-\frac{1}{2}\mid x/\lambda\mid^{\nu})}{\lambda2^{(1+1/\nu)}\Gamma(1/\nu)},$$
where $$\lambda = \sqrt{2^{-2/\nu}\frac{\Gamma(1/\nu)}{\Gamma(3/\nu)}},$$
$\ -\infty < x < \infty$, $\ 0 < \nu \le \infty$ and $\Gamma(\cdot)$ is the gamma function. When the degree of freedom $\nu$ is equal to $2$, the GED is exactly the same as the Gaussian distribution. However, the tails of GED are fatter than Gaussian distributions when $\nu < 2$, which is useful in analyzing the estimation results for leptokurtic distributions. The heavy-tailed error distributions mimic the characteristics of financial data, hence particularly relevant in judging the performance of SMLE based algorithms in this context. The Gaussian error distribution allows us to measure the performance of the change-point detection algorithm based on SMLE against Gaussian QMLE, where we expect Gaussian QMLE to be more efficient. 

Figure \ref{singlecp_plot} displays nine boxplots given different combinations of change-point positions, sample sizes, and error distributions. The vertical axis on the left denotes the estimated change-point positions, while the labels $q=(1/2, 1/3, 2/3)$ on the right are the true designed positions. As an example, for the plots in the first row marked 1/2, we expect the median of boxplots to be around $0.5$. As the sample size $n$ increases, both Gaussian QMLE and SMLE perform better, rendering accurate change-point estimations with smaller variability. Interestingly, when the true conditional distribution is Gaussian, SMLE still outperforms the Gaussian QMLE for all the sample sizes with a minor superiority. When the actual error distributions are t-distribution and GED, the SMLE has a significant advantage over Gaussian QMLE, particularly for larger sample sizes, both in terms of the average (median) and the variability of the estimated change-point. This conclusion remains consistent regardless of the position of change-points in the time series.

\begin{table}[h]
	\begin{center}
		\begin{minipage}{\textwidth}
			\caption{Bias and variance by calculating distance between estimated and true position in single change-point detection.}\label{singlecp_table}
			\begin{tabular*}{\textwidth}{@{\extracolsep{\fill}}lccccc@{\extracolsep{\fill}}}
				\toprule%
				 & Size & \multicolumn{2}{@{}c@{}}{QMLE} & \multicolumn{2}{@{}c@{}}{SMLE} \\
				\cmidrule(r){3-4}  \cmidrule(r){5-6}
				&  & Bias & Variance & Bias  & Variance\\ 
				\midrule
				Gaussian &   1000 & 0.03127 & 0.01649 &  0.00523 & 0.01106\\  
				&   2000 & 0.00728 & 0.00831 & 0.00212 & 0.00364\\ 
				&   5000 & 0.01379 & 0.00354 & 0.00265 & 0.00009\\
				\midrule
				GED &   1000 & 0.02250 & 0.01635 & -0.00353 & 0.01065\\ 
				&   2000 &  0.01597 & 0.01091  & -0.00288 & 0.00614 \\
				&   5000 & 0.01388 & 0.00611 & 0.00003 & 0.00058 \\
				\midrule
				t-distribution &   1000  & 0.00697 & 0.01555 & -0.00517 & 0.01244 \\ 
				&   2000 & 0.01058 & 0.01387 & -0.00728 & 0.00565 \\ 
				&   5000 & 0.00841 & 0.00738  & 0.00031 & 0.00096 \\ 
				\botrule
			\end{tabular*}
		\end{minipage}
	\end{center}
\end{table}

We also measured the approximate bias and variance of the estimators. As shown in table \ref{singlecp_table}, both bias and variance decrease as the sample size increases. When the error follows t-distribution and GED, SMLE performs much better with smaller bias and variance than Gaussian QMLE. Even when the error distribution is Gaussian, the performance of the SMLE is still superior to the Gaussian QMLE. This can be explained by the fact that SMLE is based on a data-dependent error distribution leading to a more precise characterization of the likelihood for a given data than the QMLE, resulting in a more robust estimation of volatility change-point.

\subsection{Identification and estimation of multiple change-points}\label{multicp}
We will now consider a more realistic scenario where there are potentially multiple change-points in volatility, and the actual number of change-points and their positions are unknown.  We will evaluate the performance of SMLE and QMLE in detecting both the number and the location of change-points. For $k$ change-points, there are $(k+1)$ GARCH(1,1) DGP with $3\times(k+1)$ parameters in the model. For the sake of brevity, in this paper, we will discuss the simulation results for $k=2$ as the conclusions are similar for larger values of $k$. We choose the first two DGP the same as the simulation setting in Section \ref{singlecp}, and the third DGP as $$\textbf{DGP 3} \ GARCH(1,1): \omega_3 = 0.2, \alpha_3 = 0.075, \beta_3 = 0.85.$$
We should note that all three DGPs used in the simulations have small $\alpha$ and large $\beta$, consistent with estimated GARCH(1,1) parameters commonly observed in financial data.  In addition, the choices of the DGPs ensure that the differences in the GARCH parameters around change-points are very subtle, making the task of change-point detection particularly challenging. We choose the length of time series for each of the three segments as $n_{DGP1} = 1000, n_{DGP2} = 500, n_{DGP3} = 500$. Therefore, the total sample size is $n = 2000$ with two change-points at the locations 1000 and 1500. The choices of true error density $f$ are the same as Section \ref{singlecp}, which are Gaussian, GED, and Student’s t-distribution. The results are based on repeated simulations of size $B = 500$ for each case. The change-points are estimated by Binary Segmentation Algorithm \ref{bsalgorithm} with SIC penalty as explained in the previous section.

\begin{figure}
	\centering
	\includegraphics[width=0.8\textwidth]{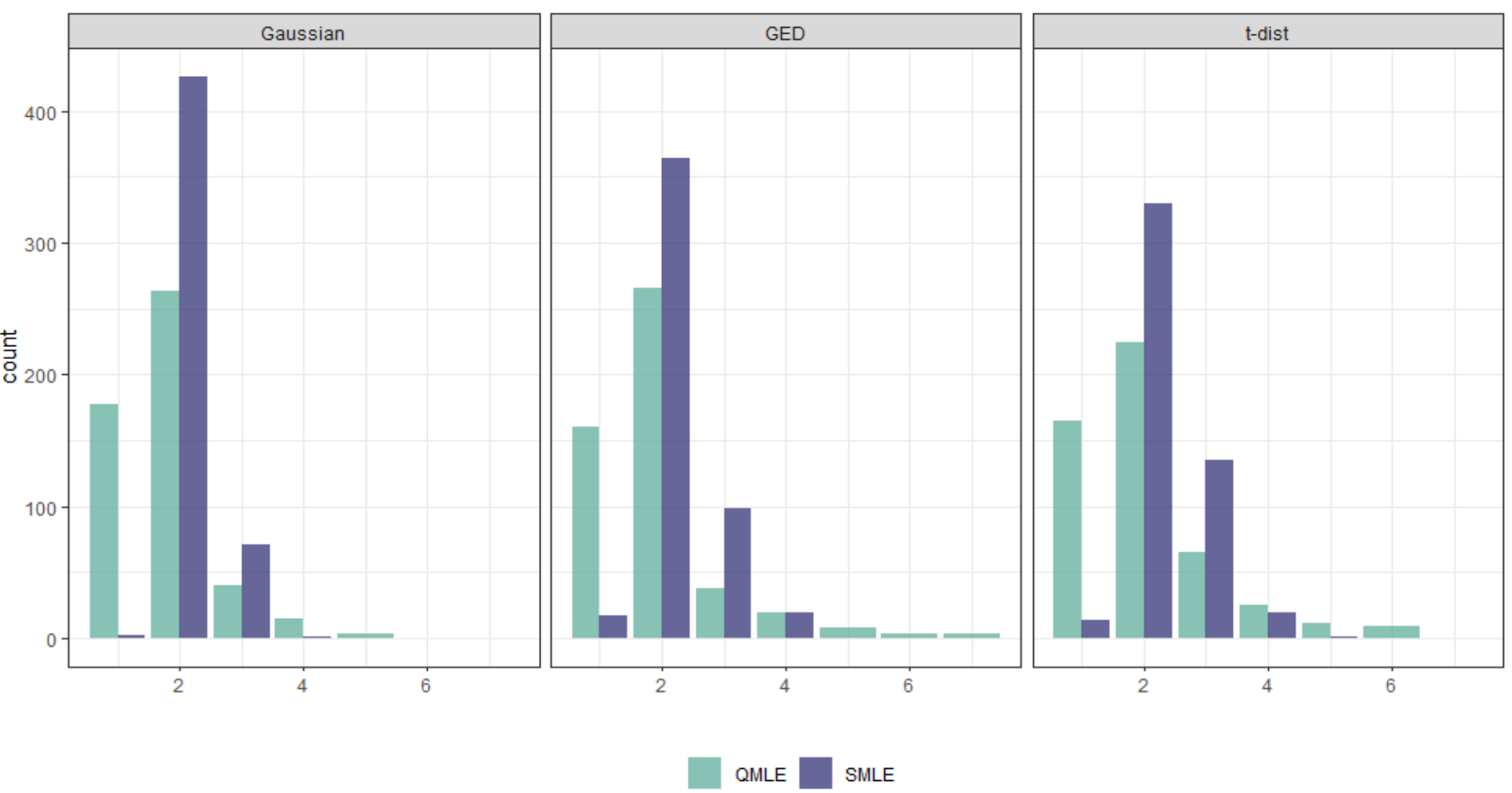}
	\caption{Bar plots of the number of estimated change-points $k$ under conditional distributions Gaussian, GED and t-distribution, where the true number of change-points is 2 and the true change-points positions are 1000 and 1500.} \label{density} 
\end{figure}

Figure \ref{density} shows the number of detected change-points based on Gaussian GARCH and SMLE models. It should be clear that under all types of conditional error distributions, the number of change-points identified by the semiparametric method is concentrated around the actual value $2$ and has significantly less variability than the QMLE. In addition, the QMLE tends to detect only one change-point rather than the true value of $2$, an outcome that is rare for SMLE. However, SMLE seems to have a slightly higher propensity to detect more change-points than the true value. In practical terms, we should note that, compared to the actual number of change-points, the detection of more change-points is better than fewer change-points. When the error distribution is Gaussian,  SMLE still outperforms Gaussian QMLE, as the frequency of correct identification of the number of change-points for SMLE is about $48\%$ greater than that of the Gaussian QMLE. It appears that the identification of change-points is particularly challenging for fat-tailed distribution. However, it is worth noting that in this case, the QMLE performs worse, frequently detecting $5$ or $6$ change-points, whereas the results based on SMLE are relatively consistent with the detected number of change-points, for the most part, is $2$ or $3$. Therefore, the simulation in Figure \ref{density} shows the superiority of SMLE over QMLE in identifying the number of change-points.

\begin{figure}
	\centering
	\includegraphics[width=1\textwidth]{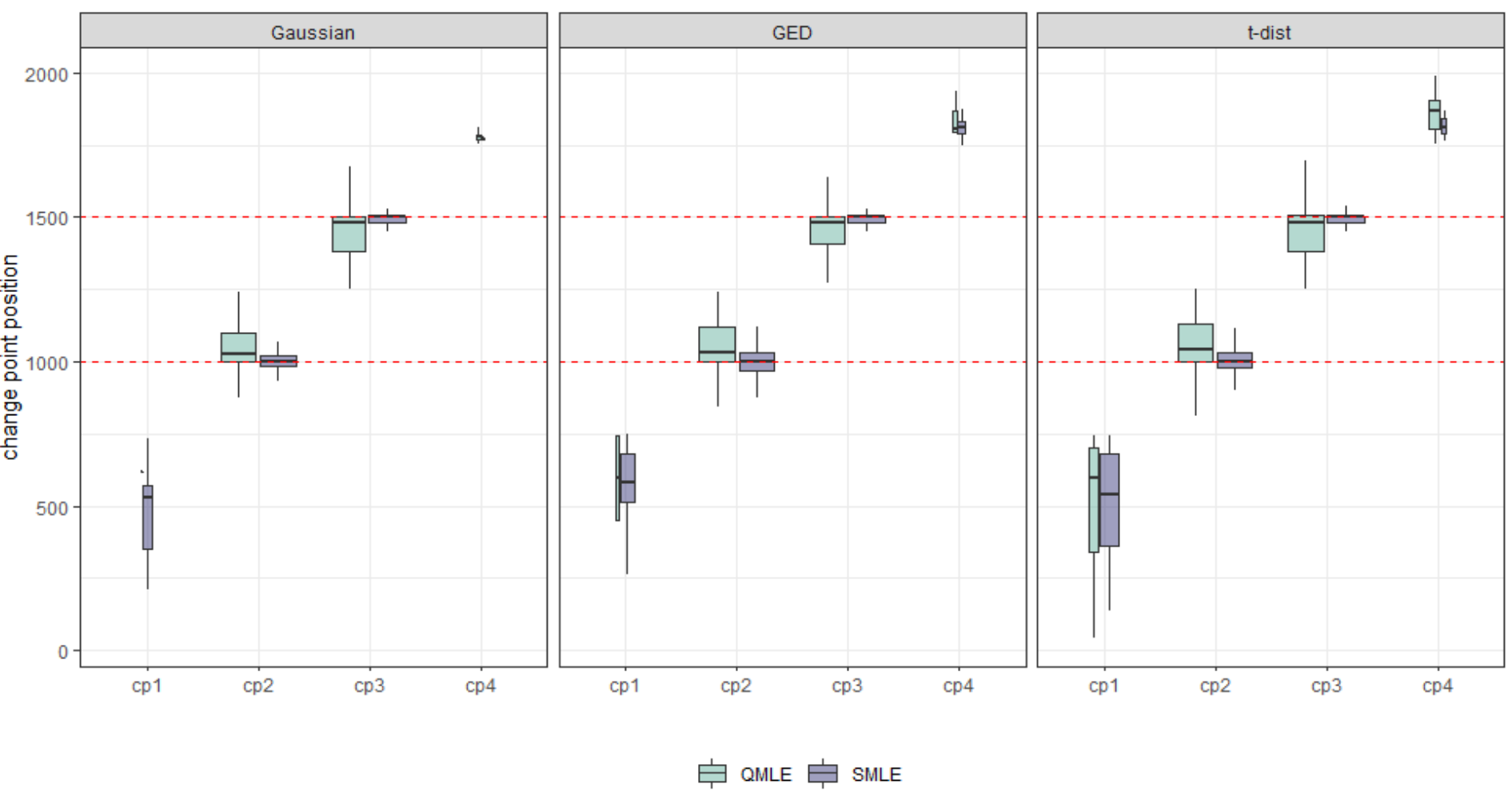}
	\caption{Boxplots of estimated locations of change-points in multiple change-points simulation. The red dashed lines indicate the true locations where the true change-points positions are 1000 and 1500.} \label{multiple} 
\end{figure}

In Figure \ref{multiple}, we illustrate the performance of the change-point detection algorithm to identify the location of the change-points correctly. Even though in our simulations, the actual number of change-points is $2$, the number of change-points identified by the algorithms depending on the simulated data can be different from $2$. Therefore, we have utilized the following approach to make comparisons between the change-point identification algorithms more informative. We denote the results of the detected change-points as \textit{cp1, cp2, cp3, cp4}, which identifies the location of the change-points in the intervals 0-750, 750 - 1250, 1250 - 1750, and $>1750$, respectively. Since the true change-points in the simulation are 1000 and 1500, most detected change-points should cluster around $cp2$ and $cp3$. The width of a boxplot is proportional to the frequency of detected change-points within a given interval. Thus, as expected, the width of the boxplot in $cp1$ and $cp4$ are narrower than that of $cp2$ and $cp3$. The dotted lines at 1000 and 1500 are the locations of the true change-points. It is clear from the plots that the SMLE-based algorithm has significantly superior performance than the Gaussian QMLE in detecting the position of the change-points. The results of the simulations for SMLE, compared to QMLE, are much tighter near the true values with substantially less variability, which remains true even when the true error distribution is Gaussian. The SMLE based algorithm is also less likely to identify spurious change-points at \textit{cp1} and \textit{cp4}. Therefore, SMLE achieves better results in multiple change-point estimations with variability reduction and enhanced accuracy.

\begin{table}[h]
	\begin{center}
		\begin{minipage}{\textwidth}
			\caption{A comparison of accuracies of change-point detection methods based on Gaussian QMLE and SMLE.}\label{multiple_error}
			\begin{tabular*}{\textwidth}{@{\extracolsep{\fill}}lccccccc@{\extracolsep{\fill}}}
				\toprule%
				 &  \multicolumn{3}{@{}c@{}}{QMLE} & \multicolumn{3}{@{}c@{}}{SMLE} \\
				\cmidrule(r){2-4}  \cmidrule(r){5-7}
				& accuracy1 &  accuracy2 & accuracy3 & accuracy1 & accuracy2 & accuracy3\\ 
				\midrule
				Gaussian & 29.0\% & 38.7\% & 49.7\% & 53.9\% & 65.7\% & 81.2\%\\ 
				GED & 29.2\% & 37.5\% & 50.5\% & 44.2\% & 57.7\% & 71.6\%\\ 
				t-dist & 23.4\% & 31.1\% & 44.0\% & 40.9\% & 55.5\% & 74.1\%\\
				\botrule
			\end{tabular*}
		\footnotetext{The table shows the percentage of change-points estimates within 10 (accuracy1), 20 (accuracy2), and 50 (accuracy3) points of the true change-points.}
		\end{minipage}
	\end{center}
\end{table}

Another way to judge the success of a change-point detection algorithm is to compare the proportion of detected change-points within a small number $m$ of the true value.  For illustration, we have considered three different values of $m$, namely, $m=10, 25, 50$, and corresponding proportion of change-points detected within $\pm m$ interval of the true values at positions $1000$ and $1500$ are labeled as \textit{accuracy1 (m=10), accuracy2 (m=25)}, and \textit{accuracy3 (m=50)}. Table \ref{multiple_error} establishes that the accuracy of the algorithm based on SMLE is significantly higher than the algorithm based on Gaussian QMLE. Even under the Gaussian error distribution, the narrowest interval with $m=10$ (\textit{accuracy1}), the SMLE-based algorithm is successful in identifying the change-points within the interval about 53.9\% of the repetitions compared to about 29.0\% for the Gaussian QMLE.

Although it has been shown in Section \ref{singlecp} that the positions of change-points do not influence the detection accuracy, we illustrate this point again in the scenario of multiple change-points. In this simulation, we set sample sizes of three DGPs to $n_{DGP1} = 500$, $n_{DGP2} = 500$, and $n_{DGP3} = 1000$, resulting in two change-points at the locations of $500$ and $1000$. All other settings are the same as before, except for the number of repeated trials chosen as $B=300$.

\begin{figure}
	\centering
	\includegraphics[width=0.8\textwidth]{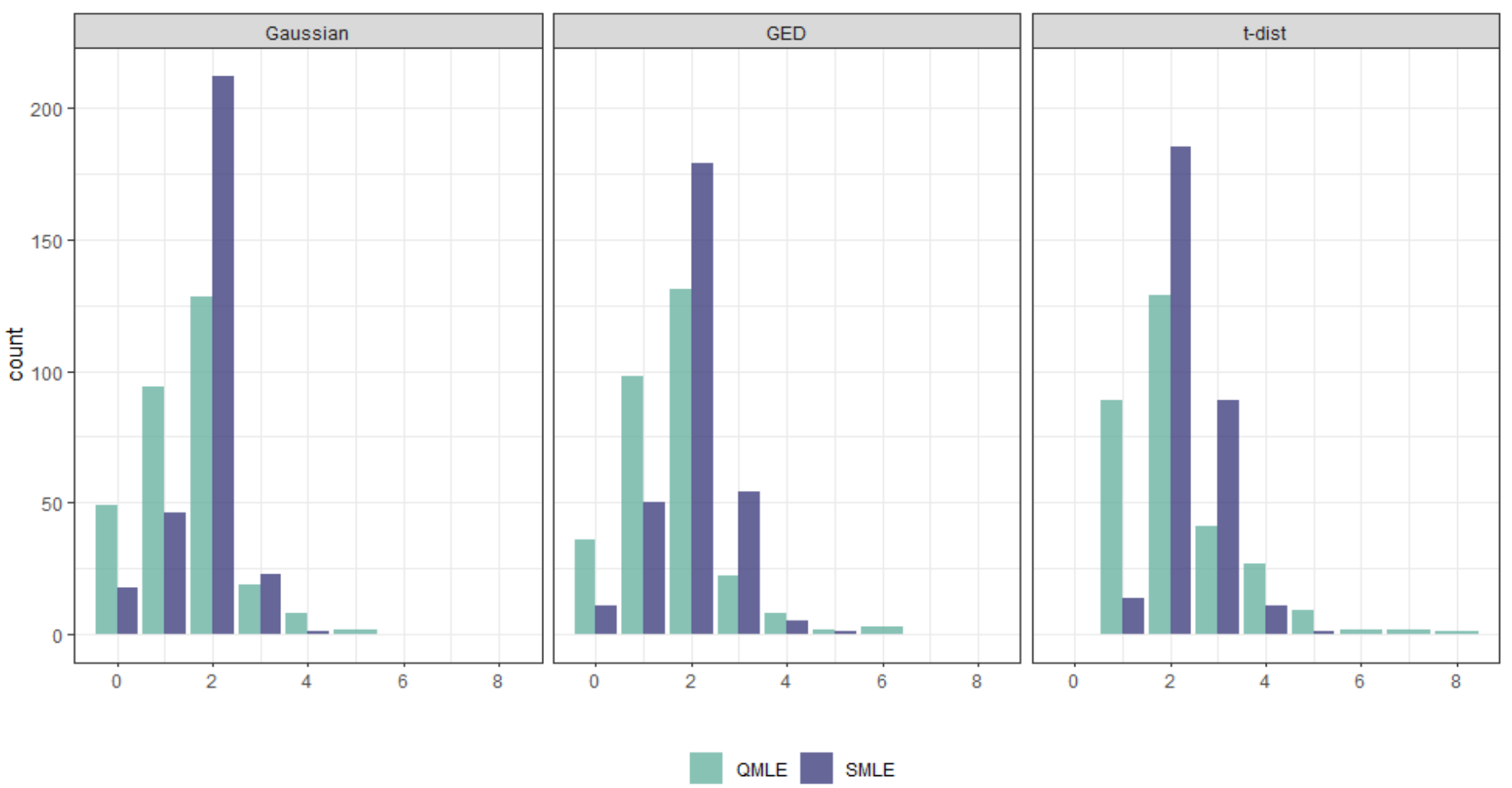}
	\caption{Bar plots of the number of estimated change-points $k$ under conditional distributions Gaussian, GED and t-distribution where the true number of change-points is 2 and the true change-points positions are 500 and 1000.} \label{density_2} 
\end{figure}

\begin{figure}
	\centering
	\includegraphics[width=1\textwidth]{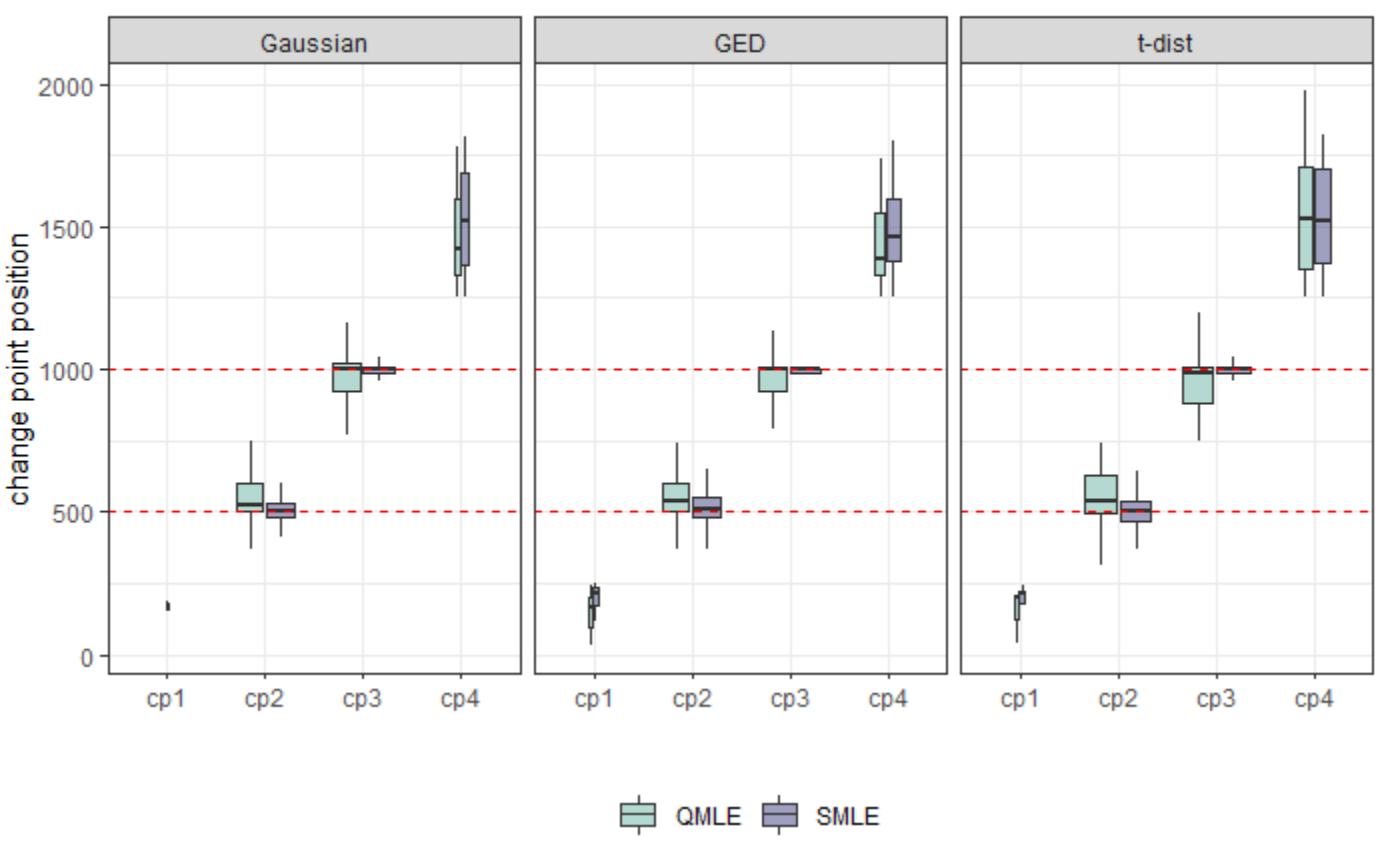}
	\caption{Boxplots of estimated locations of change-points in multiple change-points simulation. The red dashed lines indicate the true locations where the true change-points positions are 500 and 1000.} \label{multiple_2} 
\end{figure}

The resulting number of detected change-points based on Gaussian GARCH and SMLE models in the simulation are shown in Figure \ref{density_2}. The plot shows that the SMLE-based algorithm is superior to QMLE as the frequency of detecting the correct number of change-points is significantly higher for the SMLE.  In addition, Figure \ref{multiple_2} shows that the performance of the change-point detection algorithm in identifying the location of the change-points is similar to our previous observation. The boxplots of the detected change-points in the intervals \textit{cp1, cp2, cp3, cp4} defined earlier again show that SMLE is superior in multiple change-point estimations with smaller variability and increased accuracy regardless of the location of the true change-points in the time series. 

\begin{figure}
	\centering
	\includegraphics[width=0.6\textwidth]{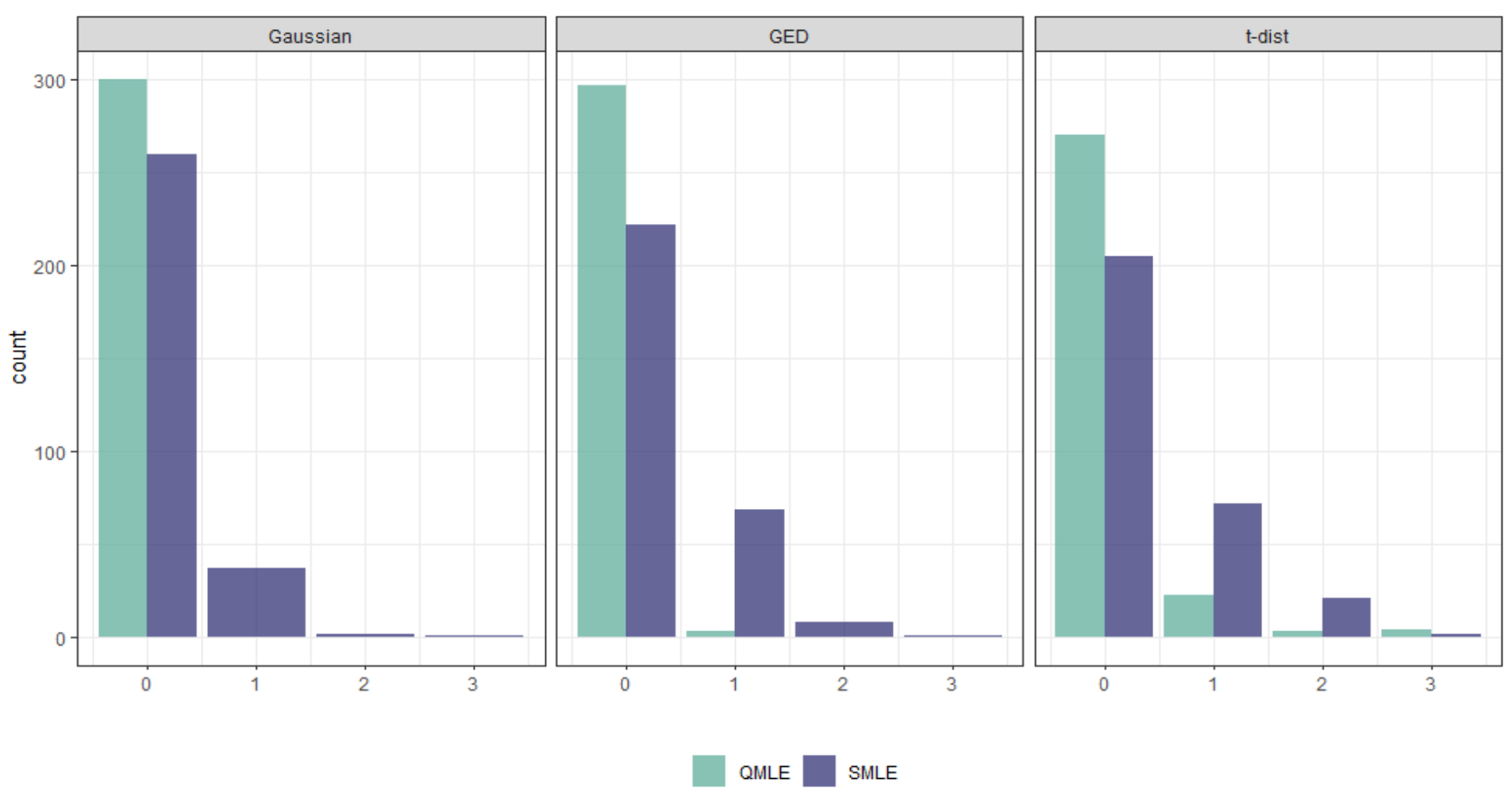}
	\caption{Bar plots of the number of estimated change-points $k$ under conditional distributions Gaussian, GED and t-distribution where no change-points exist.} \label{density_no} 
\end{figure}

Finally, for completeness, we compare the performances of the two change-point detection algorithms when there is no change-point in the simulated data. Therefore, we expect the number of detected change-points to be $0$.  We design the data generating process without any abrupt parameter change under conditional distributions Gaussian, GED, and t-distribution, with DGP setting as $$GARCH(1,1): \omega = 0.1, \alpha = 0.05, \beta = 0.9$$ and the length of the time series equal to 1000.  Figure \ref{density_no} shows the bar plots of the number of estimated change-points with $B = 300$.  The number of detected change-points concentrates on zero in both QMLE and SMLE, although SMLE detects slightly more change-points than QMLE. In conclusion, our simulations demonstrate that the performance of the semiparametric algorithm is more satisfactory than the parametric method, with high accuracy and less variability in change-point detection.

\subsection{Comparison with asymmetric GARCH models} \label{asymcp}
There are three main stylized features of financial data: volatility clustering (large changes tend to be followed by large changes), leptokurtosis (distribution of returns is fat-tailed), and the leverage effect (change in return is negatively correlated with change in volatility, hence change in return after negative shocks tend to be bigger than that after positive shocks of the same magnitude) \cite{black1976studies}. The GARCH model usually takes care of the first two features of financial data quite well but fails to model the leverage effect due to the underlying symmetry of the model structure. The development of nonlinear extensions of GARCH has been an active area of research for the last couple of decades. Among the most well-known models are the Exponential GARCH (EGARCH) model \cite{nelson1991conditional} and the GJR-GARCH model \cite{glosten1993relation}.  Other notable asymmetric GARCH models include Threshold GARCH (TGARCH) \cite{Zak.94}, Power models \cite{Eng.86}, and the nonlinear class of ARCH  models \cite{Hig.92}. \par

In this subsection, we will briefly compare the performances of change-point detection methods based on SMLE and the two most common asymmetric GARCH models, namely, EGARCH and GJR-GARCH. Exponential GARCH (EGARCH) model estimates the volatility $\sigma_t^2$ by multiplicative function of lagged error terms. In particular, EGARCH(1,1) model is given by  $$
y_t = \sigma_t z_t, \quad log(\sigma_t^2) = \omega + \{\alpha \mid z_{t - 1}\mid - E(\mid z_{t - 1}\mid) + \gamma z_{t - 1}\} + \beta\sigma^2_{t-1} $$
and GJR-GARCH(1,1) model is defined as 
$$
y_t = \sigma_t\epsilon_t, \quad \sigma_t^2 = \omega + (\alpha + \gamma I_{t - 1})y^2_{t-1}+ \beta\sigma^2_{t-1}
$$ where $$I_{t - 1} =
\begin{cases}
	0 & \text{if $y_{t - 1} \ge 0$}\\
	1 & \text{if $y_{t - 1} < 0$}\\
\end{cases}$$
and $\omega, \alpha, \beta, \gamma \ge 0$. Assumptions of $\epsilon_t \stackrel{i.i.d}{\sim} f$ and $\sigma_t$ are the same as the standard GARCH(1,1) model.

\begin{figure}
	\centering
	\includegraphics[width=0.8\textwidth]{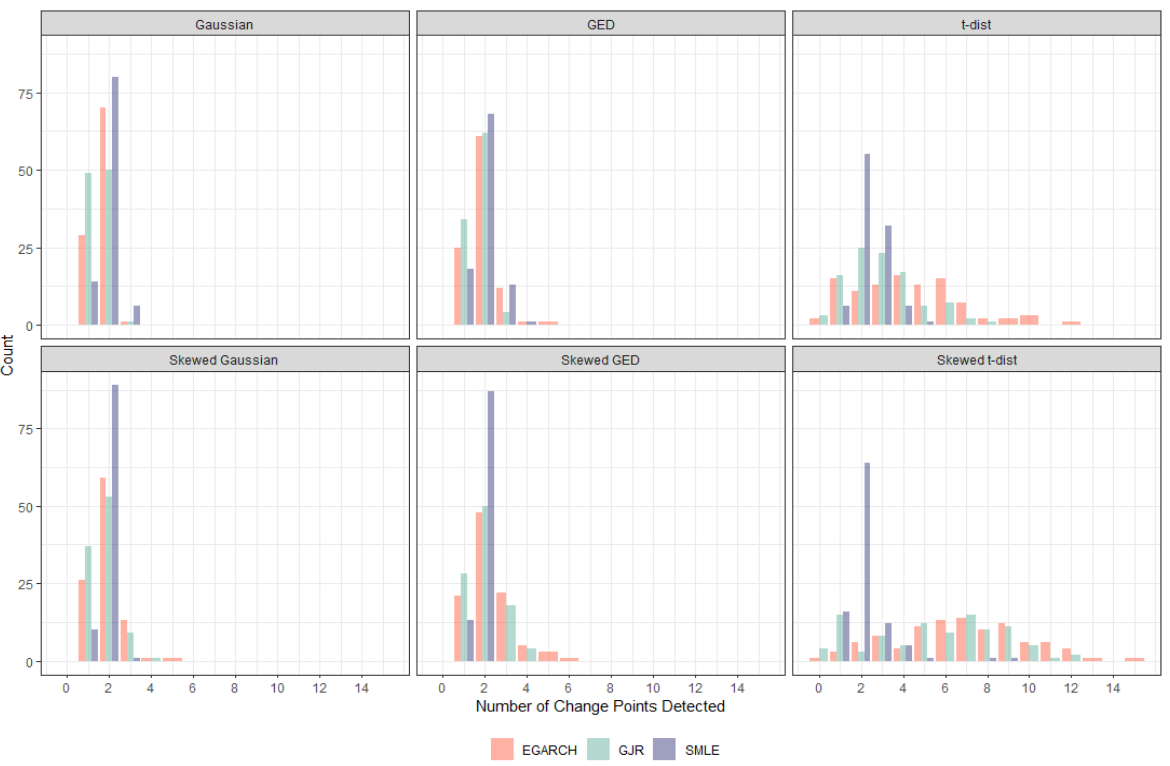}
	\caption{Barplots of the number of estimated change-points $k$ under conditional distributions Gaussian, GED,  t-distribution, skewed Gaussian, skewed GED and skewed t-distribution.  The true number of change-points in the simulation is 2 and the sample size is 2000.} \label{other_garch_bar} 
\end{figure}

We will apply these asymmetric GARCH models to identify multiple change-points and compare their performances with the SMLE-based procedure under six different error distributions. Apart from Gaussian, GED, and Student's t-distribution utilized in Section \ref{multicp}, we also consider skewed Gaussian, skewed GED, and skewed t-distributions, with the skewness parameter for all three distributions equal to 4, the resulting error distributions are right-skewed. The DGPs are identical to those in Section \ref{multicp}, including the positions of change-points and sample sizes. For the SIC penalty $\beta = p\,log(n)$ in binary segmentation, the number of parameters $p$ in EGARCH and GJR-GARCH models is 4 ($\omega, \alpha, \beta, \gamma$), which is different from Section \ref{multicp}. The results are based on repeated simulations of size $B = 100$ for each scenario.

Figure \ref{other_garch_bar} illustrates the bar plots of the number of detected change-points based on EGARCH, GJR-GARCH and SMLE when the true value is 2. Under Gaussian and GED distributions, the performances of the three methods are essentially comparable. However, under Student t-distribution and all skewed distributions, SMLE constantly outperforms EGARCH and GJR-GARCH in the sense that SMLE has by far the highest frequency of correct identification of the number of change-points. In particular, when the error distribution is given by the skewed t-distribution, the result clearly shows the vast superiority of the SMLE based change-point detection method, as the corresponding estimates of the number of change-points are overwhelmingly equal to 2, i.e., the true number of change-points, whereas the EGARCH and GJR detect a significant number of spurious change-points.

\begin{figure}
	\centering
	\includegraphics[width=1\textwidth]{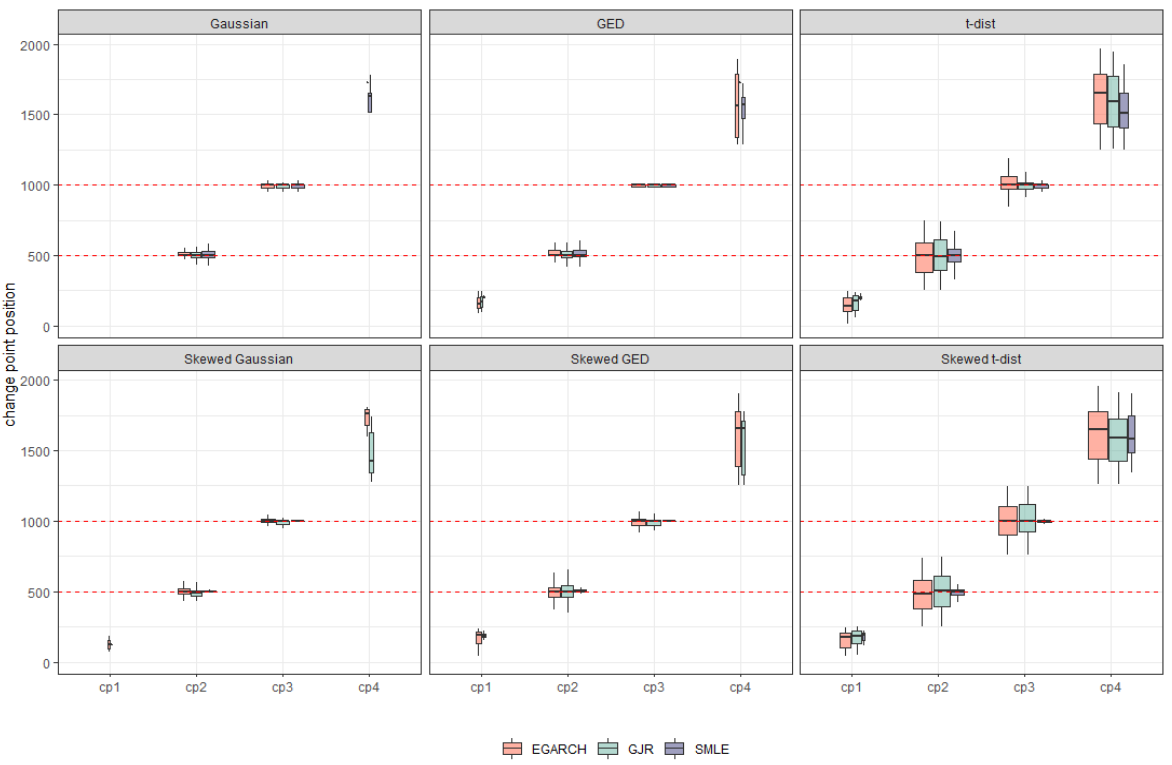}
	\caption{Boxplots of multiple change-point detection based on EGARCH, GJR-GARCH and SMLE. The true locations of the true change-points at 500 and 1000 in a sample of size 2000, and are indicated with red dashed lines.} \label{other_garch_boxplot} 
\end{figure}

Figure \ref{other_garch_boxplot} shows the multiple change-points detection results estimated by EGARCH, GJR-GARCH, and SMLE under six different residual distributions. The notations $cp1 - cp4$ are the same as that in Section \ref{multicp}, and most of detected change-points are around $cp2$ and $cp3$, as expected. In particular, the medians of all three approaches in $cp2$ and $cp3$ are close to the true change-points at 500 and 1000 marked with red dashed lines. Under the Gaussian and GED distributions, the difference in performance among the three methods is negligible. However, under the t-distribution and skewed t-distribution, the variability of the EGARCH and GJR estimates is significantly higher than that of the SMLE, which leads to more robust and stable change-point detection results. Furthermore, the frequency of change-point detection based on SMLE characterized by the location and width of the box plots at incorrect locations $cp1$ and $cp4$ are significantly rarer than EGARCH and GJR-GARCH.  For example, there are almost no detected change-points in $cp1$ and $cp4$ via the SMLE approach under the skewed Gaussian, and skewed GED cases, but the existence of detected change-points based on EGARCH and GJR-GARCH is evident. Therefore, the SMLE approach is not only better than the standard QMLE Gaussian GARCH model as shown in Section \ref{multicp}, but also performs better than other asymmetric GARCH models in change-point detection procedures.

\subsection{Application to financial time series}\label{application}
In this section, we explore the performance of the semiparametric change-point detection algorithm in identifying structural changes in volatility of financial time series.  In particular, we consider the time series on the daily open values of S \& P 500 Index and Dow Jones Industrial Average from June 2015 to June 2020 ($n = 1295$), both are prominent stock indices measuring global financial market performance. S \& P 500 tracks 500 publicly traded companies on stock exchanges and Dow Jones is based on the stock prices of 30 large companies representing their respective industries in the United States. As a consequence of different stock compositions and calculation methods, the performance of these benchmarks differs. Although the general trends of the two series are similar, the magnitudes of change over time can be distinct.

We employ the binary segmentation algorithm based on semiparametric GARCH models to split the recent five-year data into segments. In the real world application, it is difficult to prove whether the detected change-points are the true reflections of the changes in the data generating process. However, it may be possible to explore the association between certain events and corresponding structural changes in the volatility of financial time series. Thus, our aim is to find events corresponding to these change-points, which we use as surrogates to the change-point identification accuracy. If the change-points appear at close proximity for both stock indices, it also demonstrates the credibility of change-point detection algorithm.

\begin{figure}
	\centering
	\includegraphics[width=1\textwidth]{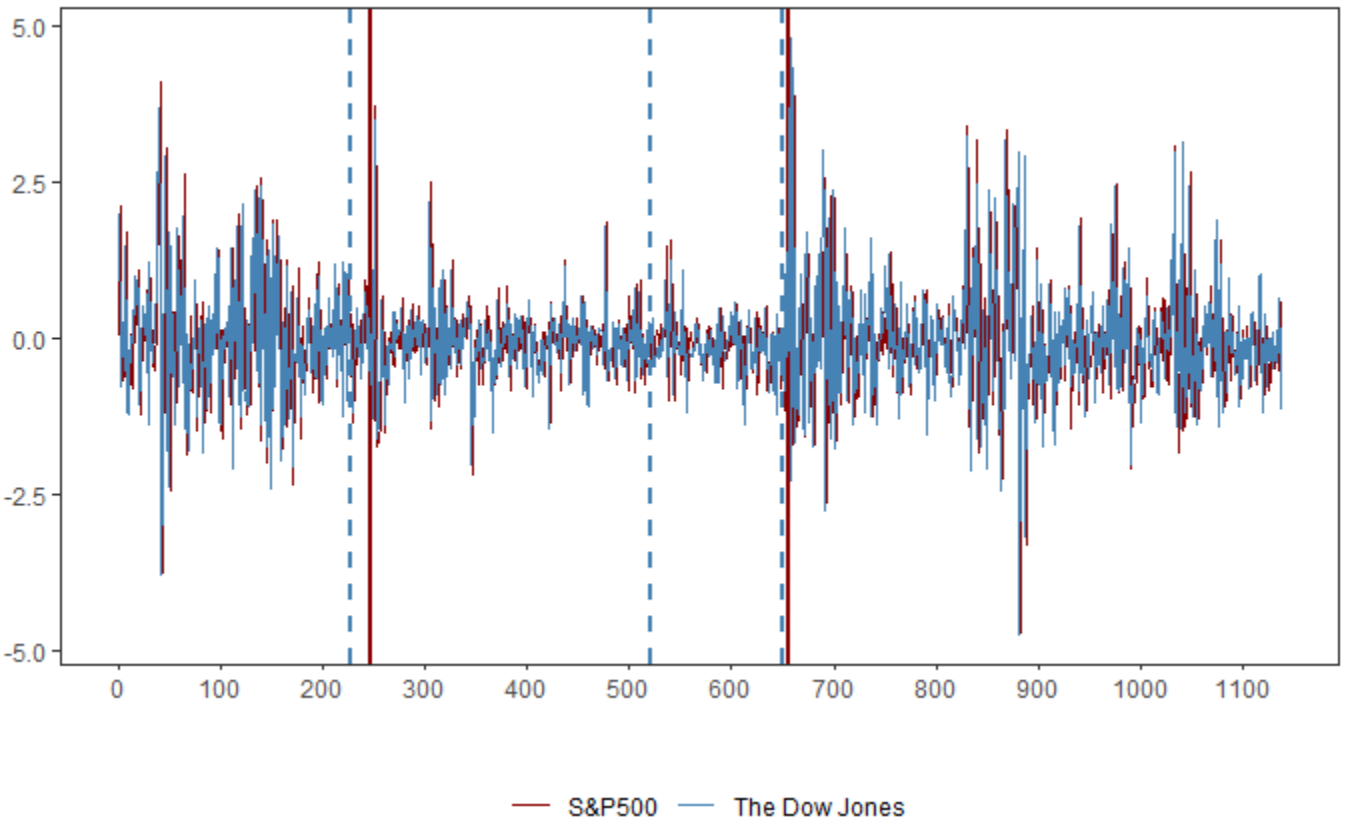}
	\caption{Location of estimated change-points of S\&P 500 and Dow Jones Indices from June 2015 to June 2020 indicated by blue and red lines, respectively.} \label{app} 
\end{figure}

Figure \ref{app} shows the positions of estimated change-points by SMLE on the S \& P 500 and Dow Jones indices returns indicated by red and blue lines, respectively. This data set covers the period from June 26, 2015, to December 31, 2019, and the data length is 1138. The number of detected change-points for S \& P 500 and Dow Jones are 2 and 3, respectively, and the locations of the change-points of S \& P 500 are 247 and 655, whereas the same for Dow Jones are 227, 521, and 649. We can see from this plot that the locations of change-points on these two indices overlap greatly. These estimated locations of abrupt changes are indeed related to real events. The first change-point that $\hat\tau_{S \& P 500}^1 = 247$ and $\hat\tau_{Dow Jones}^1 = 227$ is around June 2016 (Brexit) and the last change-point $\hat\tau_{S \& P 500}^2 = 655$ and $\hat\tau_{Dow Jones}^3= 649$ is near February 2018 (US stock market crash). These observations constitute indirect evidence of the validity of the change-point detection algorithm based on the semiparametric GARCH model introduced in this paper. 

%Figure \ref{app} shows the positions of estimated change-points on the S \& P 500 and Dow Jones indices indicated by red and blue lines, respectively. The number of detected change-points for the S \& P 500 and Dow Jones is the same as $2$. In addition, the locations of change-points on these two indices overlap to a great extent. These estimated abrupt changes are indeed related to real events, such as $\hat \tau = 250$ (June 2016) Brexit and $\hat \tau = 652$ (February 2018) US stock market crash. These observations constitute indirect evidence of the validity of the change-point detection algorithm based on the semiparametric GARCH model introduced in this paper.

\section{Discussion and conclusion}\label{Section5}
In this paper, we have explored the problem of the detection of volatility change-points in financial data.  What sets the investigation presented in this paper apart from comparable work in the literature is that the estimation is carried out without any prior parametric assumption on the condition error distribution. This is a semiparametric approach in the sense that the likelihood retains the structural advantages of the GARCH process but allows the flexibility of nonparametric conditionalal error distribution. This is a key advantage as it is rare for the conditional error distribution of a true data generating process in a financial time series to be consistent with a specific parametric distribution.

The change-point detection algorithm introduced in the paper is based on finding the optimal segments of data by minimizing a penalized cost function, where the cost function is given by the likelihood, or more specifically -2$\times$ log-likelihood, calculated from the Gaussian and semiparametric GARCH models. In general, the semiparametric estimation approaches do not pre-specify the likelihood function, and therefore it is more robust than the traditional quasi-likelihood methods. As established in this paper, via extensive simulations, this is an essential advantage in the context of change-point detection and estimation of volatility. The results show that SMLE based algorithm uniformly outperforms Gaussian QMLE regardless of the sample size, relative location of the change-points, and the true error distribution. It is particularly interesting to note that the semiparametric algorithm is superior to Gaussian QMLE even when the conditional error distribution is actually Gaussian.  This can be explained by the fact that the SMLE based algorithm offers a data-dependent approach to change-point detection and is sufficiently flexible in accommodating any nuances of the data, which is not the case with a rigid parametric framework. 

It is worth noting that the performance of the change-point detection algorithms differs depending on the true error distribution utilized in the data generating process. In particular, the performance of change-point detection algorithms is affected when the error distributions are heavy-tailed. However, we have shown that even in the case of t-distribution, the SMLE based algorithm performs significantly better than QMLE for larger sample sizes, which is reassuring as most financial data are generally quite long.  Although we did not investigate it in this paper, it may be possible to mitigate the effect of leptokurtic distributions by considering a modified version of the kernel density estimator used in the SMLE. One possibility is to consider a kernel density estimator using the Champernowne transformation that has been shown to have improved performance in the context of heavy-tailed distributions compared to a regular density kernel estimator \cite{Buch2006}. We have also investigated the change-point detection performance of the semiparametric method compared to the two most common asymmetric GARCH models, and it shows the suitability of the proposed method in detecting volatility change-points of financial data in the presence of leverage effect.

The results show that if there is no change-point, the algorithm's performance based on Gaussian QMLE is, albeit marginally, superior to the semiparametric algorithm proposed in the paper.  However, the advantage of the Semiparametric MLE is realized if there are structural changes in the data.  This is because even though the true error distribution is Gaussian,  in the presence of one or more change-points, the data is not consistent with the model, and the adaptive nature of the semiparametric algorithm has a distinct advantage over the Gaussian QMLE.  In essence, the key takeaway of the paper is that the proposed semiparametric approach shines under the model miss-specification.  A financial time series observed over a long period is likely to have structural changes. The traditional method of volatility estimation based on a single GARCH(1,1)  (or other approaches) process will not result in a reliable estimate of volatility. The proposed method mitigates the risk by adapting to the structural changes, resulting in a robust volatility estimate.

One interesting question for future research is to study the theoretical properties of the change-point estimator based on the SMLE algorithm.  It is a challenging question, and as far as we know, no work has been done in the literature in this vein even in the QMLE framework. The asymptotic properties of the parameter estimates of the GARCH(1,1) process under semiparametric likelihood have been studied in detail by Di and Gangopadhyay \cite{di2011efficiency, di2013one}.   Since the estimators are derived from kernel density estimates, the semiparametric GARCH estimators are inherently biased, and the bias goes to $0$ very slowly for large  $n$. This creates additional technical hurdles in developing the theoretical properties of the change-point estimator based on the SMLE algorithm.

In summary, the paper addresses the important question of identification of volatility change-points in a financial time series and introduces a novel semiparametric approach to change-point detection.  The extensive simulations presented in the paper clearly establish the superiority of the semiparametric approach over the traditional Gaussian GARCH model in the context of change-point detection. The ideas and methodologies presented in the paper open up a new avenue of further research on the important question of volatility change-point in financial data.

%\section*{Acknowledgement}
%We are indebted to an anonymous reviewer and an editor for providing insightful comments and directions %for additional work that has significantly improved the paper.


\bibliography{sn-bibliography}% common bib file
%\input sn-article.bbl


\end{document}